\input harvmac
%
%
\noblackbox

\newcount\figno
\figno=0
\def\fig#1#2#3{
\par\begingroup\parindent=0pt\leftskip=1cm\rightskip=1cm\parindent=0pt
\baselineskip=11pt
\global\advance\figno by 1
\midinsert
\epsfxsize=#3
\centerline{\epsfbox{#2}}
\vskip 12pt
{\bf Fig.\ \the\figno: } #1\par
\endinsert\endgroup\par
}
\def\figlabel#1{\xdef#1{\the\figno}}
\def\encadremath#1{\vbox{\hrule\hbox{\vrule\kern8pt\vbox{\kern8pt
\hbox{$\displaystyle #1$}\kern8pt}
\kern8pt\vrule}\hrule}}

\def\frac#1#2{{#1 \over #2}}

\def\p{\partial}
\def\semi{\subset\kern-1em\times\;}
\def\bar#1{\overline{#1}}
\def\sqr#1#2{{\vcenter{\vbox{\hrule height.#2pt 
\hbox{\vrule width.#2pt height#1pt \kern#1pt \vrule width.#2pt}
\hrule height.#2pt}}}}

\def\p{\partial}

\def\ad{\bar a}

%

%
\def\ap{\alpha'}
\def\p{\partial}
\def\zb{\bar{z}}
\def\Xt{\tilde{X}}
\Title{\vbox{\baselineskip12pt
\hbox{hep-th/0206080}
\hbox{UCLA-02-TEP-13}
\vskip-.5in}}
{\vbox{\centerline{Strings in Noncompact Spacetimes: }
\smallskip
\centerline{Boundary Terms and Conserved Charges}}}
\centerline{Per Kraus, Anton Ryzhov, and  Masaki Shigemori}
\bigskip\medskip
\centerline{\it Department of Physics and Astronomy,}
\centerline{\it  UCLA, Los Angeles, CA 90095-1547,} 
\centerline{\tt pkraus, ryzhovav, shige@physics.ucla.edu}
\medskip
\baselineskip18pt
\medskip\bigskip\medskip\bigskip\medskip
\baselineskip16pt
\noindent

We study some of the novel properties of conformal field theories with
noncompact target spaces as applied to string theory.  Standard CFT 
results get corrected by boundary terms in the target space in a way
consistent with the expected spacetime physics.  For instance,
one-point functions of general operators on the sphere and boundary operators
on the disk need not vanish; we show that they are instead 
equal to boundary terms 
in spacetime.  By applying this result to vertex operators for
spacetime gauge transformations with support at infinity, we derive 
formulas for conserved gauge charges in string theory.  This 
approach provides a direct
CFT definition of ADM energy-momentum in string theory.

\Date{June, 2002}
\lref\Polch{
J. Polchinski,
``String Theory,''
Cambridge University Press (1998).
}

\lref\TaniiUG{
Y.~Tanii and Y.~Watabiki,
Nucl.\ Phys.\ B {\bf 316}, 171 (1989).
}

\lref\WeisbergerQD{
W.~I.~Weisberger,
Nucl.\ Phys.\ B {\bf 284}, 171 (1987).
}

\lref\DouglasEU{
M.~R.~Douglas and B.~Grinstein,
Phys.\ Lett.\ B {\bf 183}, 52 (1987)
[Erratum-ibid.\  {\bf 187B}, 442 (1987)].
}

\lref\HarveyWM{
J.~A.~Harvey, D.~Kutasov, E.~J.~Martinec and G.~Moore,
arXiv:hep-th/0111154.
}

\lref\KutasovQP{
D.~Kutasov, M.~Marino and G.~W.~Moore,
JHEP {\bf 0010}, 045 (2000)
[arXiv:hep-th/0009148].
}

\lref\WittenQY{
E.~Witten,
Phys.\ Rev.\ D {\bf 46}, 5467 (1992)
[arXiv:hep-th/9208027];
Phys.\ Rev.\ D {\bf 47}, 3405 (1993)
[arXiv:hep-th/9210065].
}

\lref\GerasimovZP{
A.~A.~Gerasimov and S.~L.~Shatashvili,
JHEP {\bf 0010}, 034 (2000)
[arXiv:hep-th/0009103].
}

\lref\FradkinQD{
E.~S.~Fradkin and A.~A.~Tseytlin,
Phys.\ Lett.\ B {\bf 163}, 123 (1985).
}

\lref\FradkinPQ{
E.~S.~Fradkin and A.~A.~Tseytlin,
Phys.\ Lett.\ B {\bf 158}, 316 (1985).
}

\lref\GiveonNS{
A.~Giveon, D.~Kutasov and N.~Seiberg,
Adv.\ Theor.\ Math.\ Phys.\  {\bf 2}, 733 (1998)
[arXiv:hep-th/9806194];
D.~Kutasov and N.~Seiberg,
JHEP {\bf 9904}, 008 (1999)
[arXiv:hep-th/9903219].
}

\lref\feff{
C.~Fefferman and C.~.~Graham, ``Conformal Invariants'', in {\it Elie
Cartan et les Math\'{e}matiques d'aujourd'hui} (Ast\'{e}risque, 1985) 95.
} 

\lref\MaldacenaHW{
J.~M.~Maldacena and H.~Ooguri,
J.\ Math.\ Phys.\  {\bf 42}, 2929 (2001)
[arXiv:hep-th/0001053];
%
J.~M.~Maldacena, H.~Ooguri and J.~Son,
J.\ Math.\ Phys.\  {\bf 42}, 2961 (2001)
[arXiv:hep-th/0005183];
%
J.~M.~Maldacena and H.~Ooguri,
Phys.\ Rev.\ D {\bf 65}, 106006 (2002)
[arXiv:hep-th/0111180].
}

\lref\SeibergEB{
N.~Seiberg,
Prog.\ Theor.\ Phys.\ Suppl.\  {\bf 102}, 319 (1990).
}

\lref\ZamolodchikovGT{
A.~B.~Zamolodchikov,
JETP Lett.\  {\bf 43}, 730 (1986)
[Pisma Zh.\ Eksp.\ Teor.\ Fiz.\  {\bf 43}, 565 (1986)].
}

\lref\AdamsSV{
A.~Adams, J.~Polchinski and E.~Silverstein,
JHEP {\bf 0110}, 029 (2001)
[arXiv:hep-th/0108075].
}

\lref\EmparanPM{
R.~Emparan, C.~V.~Johnson and R.~C.~Myers,
Phys.\ Rev.\ D {\bf 60}, 104001 (1999)
[arXiv:hep-th/9903238].
}

\lref\AharonyTI{
O.~Aharony, S.~S.~Gubser, J.~M.~Maldacena, H.~Ooguri and Y.~Oz,
Phys.\ Rept.\  {\bf 323}, 183 (2000)
[arXiv:hep-th/9905111].
}

\lref\KrausDI{
P.~Kraus, F.~Larsen and R.~Siebelink,
Nucl.\ Phys.\ B {\bf 563}, 259 (1999)
[arXiv:hep-th/9906127].
}

\lref\mtw{C.W.~Misner, K.S.~Thorne, and J.A.~Wheeler, ``Gravitation'',
W.H. Freeman (1973).}

\lref\BalasubramanianRE{
V.~Balasubramanian and P.~Kraus,
Commun.\ Math.\ Phys.\  {\bf 208}, 413 (1999)
[arXiv:hep-th/9902121].
}

\lref\HenningsonGX{
M.~Henningson and K.~Skenderis,
JHEP {\bf 9807}, 023 (1998)
[arXiv:hep-th/9806087].
}

\lref\BrownBR{
J.~D.~Brown and J.~W.~York,
Phys.\ Rev.\ D {\bf 47}, 1407 (1993).
}

\lref\BrownNW{
J.~D.~Brown and M.~Henneaux,
Commun.\ Math.\ Phys.\  {\bf 104}, 207 (1986).
}

\lref\GibbonsUE{
G.~W.~Gibbons and S.~W.~Hawking,
Phys.\ Rev.\ D {\bf 15}, 2752 (1977).
}

\lref\WaldWA{
R.~M.~Wald and A.~Zoupas,
Phys.\ Rev.\ D {\bf 61}, 084027 (2000)
[arXiv:gr-qc/9911095].
}

\lref\PolchinskiZF{
J.~Polchinski,
Commun.\ Math.\ Phys.\  {\bf 104}, 37 (1986).
}

\lref\TseytlinTV{
A.~A.~Tseytlin,
Phys.\ Lett.\ B {\bf 208}, 221 (1988).
}

\lref\DHokerTA{
E.~D'Hoker and D.~H.~Phong,
Rev.\ Mod.\ Phys.\  {\bf 60}, 917 (1988).
}

\lref\TseytlinDJ{
A.~A.~Tseytlin,
arXiv:hep-th/9908105.
}

\lref\PolchinskiDY{
J.~Polchinski,
Nucl.\ Phys.\ B {\bf 303}, 226 (1988).
}

\lref\TeschnerFT{
J.~Teschner,
Nucl.\ Phys.\ B {\bf 546}, 390 (1999)
[arXiv:hep-th/9712256];
Nucl.\ Phys.\ B {\bf 571}, 555 (2000)
[arXiv:hep-th/9906215].
}

\lref\SusskindSM{
L.~Susskind and J.~Uglum,
Phys.\ Rev.\ D {\bf 50}, 2700 (1994)
[arXiv:hep-th/9401070].
}

\lref\KazakovPJ{
V.~A.~Kazakov and A.~A.~Tseytlin,
JHEP {\bf 0106}, 021 (2001)
[arXiv:hep-th/0104138].
}

\lref\LiuNZ{
J.~Liu and J.~Polchinski,
Phys.\ Lett.\ B {\bf 203}, 39 (1988).
}

\lref\TseytlinTV{
A.~A.~Tseytlin,
Phys.\ Lett.\ B {\bf 208}, 221 (1988);
Phys.\ Lett.\ B {\bf 194}, 63 (1987).
}

\lref\deBoerPP{
J.~de Boer, H.~Ooguri, H.~Robins and J.~Tannenhauser,
JHEP {\bf 9812}, 026 (1998)
[arXiv:hep-th/9812046].
}

\lref\GubserBC{
S.~S.~Gubser, I.~R.~Klebanov and A.~M.~Polyakov,
Phys.\ Lett.\ B {\bf 428}, 105 (1998)
[arXiv:hep-th/9802109].
}

\lref\WittenQJ{
E.~Witten,
Adv.\ Theor.\ Math.\ Phys.\  {\bf 2}, 253 (1998)
[arXiv:hep-th/9802150].
}

\lref\CallanAT{
C.~G.~Callan, J.~A.~Harvey and A.~Strominger,
arXiv:hep-th/9112030.
}

\lref\weinberg{S. Weinberg, ``Gravitation and Cosmology'', John Wiley and
Sons (1972).}


\newsec{Introduction}

Historically, the solutions of string theory attracting the most attention
have been of the form $M^4 \times K$, where $M^4$ is four dimensional
Minkowski spacetime and $K$ is some compact space.  Abstractly, $K$ is
described by a compact, unitary, (super)conformal field theory of 
appropriate central charge.  $M^4$, on the other hand, is described by a
noncompact, nonunitary free CFT.   

There are many instances in which one is interested in a more general
class of solutions, namely those corresponding to noncompact interacting
CFTs.  The study of string solitons, cosmological spacetimes and anti-de
Sitter spacetimes are just a few examples.   Our focus here will be on the
new features that arise due to the noncompactness of the underlying CFT. 

Compactness and unitarity lead to some familiar theorems which must
be reexamined in the more general context.  An example is Zamolodchikov's
c-theorem \ZamolodchikovGT, which establishes the 
existence of a function that monotonically
decreases under RG flow and is stationary with respect to marginal 
perturbations.  The proof of the theorem is based on properties of
two-point energy-momentum
 tensor correlators $\langle T_{ab} T_{cd}\rangle$, but these
correlators are generically ill-defined in a noncompact theory.   And indeed,
the theorem does not hold in the noncompact setting 
\refs{\PolchinskiDY,\AdamsSV,\HarveyWM}.

Another example ---  of direct interest to us here ---  are the conformal
Ward identities obeyed by correlators in a compact CFT.  These follow from
the operator product expansion
\eqn\ka{ 
T(z') {\cal O}(z,\bar{z}) ~\sim~ \ldots ~ 
{h {\cal O}(z,\bar{z})
\over (z' -z)^2} + {\p  {\cal O}(z,\bar{z}) \over z'-z},}
and holomorphicity of the energy-momentum tensor
\eqn\kb{ {\p \over \p \zb'} \langle T(z') {\cal O}_1(z_1,\zb_1) \ldots 
{\cal O}_n(z_n,\zb_n) \rangle = 0, \quad z' \neq z_i.}
An elementary consequence is the vanishing of one-point functions of
operators on the sphere 
\eqn\kc{\langle {\cal O}(z,\zb) \rangle_{S_2} = 0, {\quad} (h,\tilde{h})
\neq (0,0),}
and of one-point functions of boundary operators on the disk
\eqn\kd{\langle {\cal O_{{\rm bndy}}}(z,\zb) \rangle_{D_2} = 0, {\quad} h
\neq 0.}
In the noncompact case \kb\ need not be true, and with it the 
consequences \kc\ and \kd.  For example, nonzero one-point functions
appear in Liouville theory;  see, {\it e.g.}, \SeibergEB.

To understand this claim, note that in a CFT with noncompact target space
there are two classes of states and operators, distinguished by their
normalizability properties.  There does not exist a normalizable
 $SL(2)$ invariant vacuum state; normalizable states instead correspond to 
inserting operators which fall off sufficiently rapidly at infinity in the
target space.  These operators correspond, roughly speaking, to normalizable
wavefunctions on the target space.  Correlation functions of such operators
will obey the standard theorems \kb-\kd.  

However, the remaining ``nonnormalizable operators'' play a crucial 
role in noncompact theories.   In the AdS/CFT correspondence correlation
functions in the boundary theory are related to worldsheet 
CFT correlation functions
of such nonnormalizable operators 
\refs{\GubserBC,\WittenQJ}.  Therefore, these vertex 
operators are central in the study of string theory in AdS$_3$ 
\refs{\GiveonNS,\MaldacenaHW,\TeschnerFT}. 
   Since what distinguishes the
normalizable and nonnormalizable operators is their behavior at infinity,
one might expect that violations of \kb-\kd\ can be expressed as boundary
terms at infinity in the target space.   We will show that this is the 
case, giving explicit formulas for the resulting boundary integrals
in terms of CFT correlators.   For instance, in the general Weyl invariant
bosonic
nonlinear sigma model \kc\ is replaced by
\eqn\ke{\langle {\cal O}(z,\zb) \rangle_{S_2} ~ =~
-{1 \over 2 h} (V_M)^{D/2-1}\int \! d^{D-1}S_\mu \, 
 \int \! d^2z'\, (z'-z)e^{2\omega(z',\bar{z}')}\langle
\p X^\mu(z',\bar{z}'){\cal O}(z,\zb)\rangle'_{S_2}.}
Here $e^{2\omega(z',\bar{z}')}$ is the worldsheet conformal factor, $V_M$
is the volume of the worldsheet,
and $\langle \ldots \rangle'$ denotes a path integral with the constant mode
integration omitted. 

These nonzero one-point functions play an important and desirable role 
from the spacetime point of view.  Consider the spacetime action $S$ 
whose Euler-Lagrange equations reproduce the Weyl invariance conditions
of the sigma model.   The variation of $S$ with respect to the spacetime
fields $\phi_i$ is equal to a bulk term, 
which vanishes on solutions to the equations of motion, plus a boundary 
term:
\eqn\kf{\delta S = \int \! d^Dx\, \left\{ =0~ {\rm by ~ e.o.m.}\right\}
+ \int \! d^{D-1}S^\mu\, \Pi^i_\mu \delta \phi_i.}
The statement that the spacetime action varies by a boundary term is
directly related to the CFT results discussed above.  In the case of
open string theory at the level of disk amplitudes, the relation follows
from the fact that the on-shell spacetime action is proportional to 
the partition function of the corresponding CFT on the disk
\refs{\FradkinQD,\WittenQY,\GerasimovZP,\KutasovQP}.  The variation
\kf\ is then the one-point function of an operator on the boundary of 
the disk, so the 
statement that the one-point function is nonzero and equal to a boundary
term in spacetime 
is consistent with the result \kf.  In closed string theory
the connection is not quite as precise, as the spacetime action does
not seem to be proportional to the sphere partition function; we will have
more to say about this at the end of the paper.   

The boundary terms in \kf\ are of interest from several points of view.
In the AdS/CFT correspondence they define correlators of the boundary
theory, as we have already mentioned.  Another important application,
which we develop here, is
in defining conserved charges associated with gauge symmetries 
\refs{\BrownBR}.  The idea
is to consider \kf\ with $\delta \phi_i$ corresponding to a gauge 
transformation with support at infinity.  Being a gauge transformation,
we must of course have $\delta S =0$ for such a variation.  On the other hand,
after integration by parts the boundary term takes the form of a time
derivative of a charge, so one  learns that this charge is conserved.
This same procedure can be carried out in the string path integral, leading
to a direct CFT definition of conserved charges in string theory. 
These charges include electromagnetic and anti-symmetric tensor charges,
as well as those associated with the gravitational field.
In asymptotically flat spacetime the latter correspond to mass, 
momentum, and angular momentum.  For the conserved energy-momentum we find 
the result
\eqn\kg{P^\mu ~ \propto ~  \int \! d^{D-2}S^i
\int \! d^2z \int \! d^2z'\, (z'-z)e^{2\omega(z',\bar{z}')}\langle
\p X^i(z',\bar{z}')
(\p X^0 \bar{\p}X^\mu + \p X^\mu \bar{\p}X^0)(z,\zb)
 \rangle'_{S_2}.}
(We suppressed some corrections from the dilaton; see section 6.)
This will be shown to coincide with the standard ADM definition.  
As far as we know, a direct CFT definition of ADM energy and momentum
has not been given before.  

The alternative approach to deriving conserved charges in string theory
is to first derive the low energy effective action in the $\ap$ expansion,
and then to proceed as in field theory.  In asymptotically flat spacetime
the two derivative approximation to the action is usually sufficient, since 
derivatives become small at infinity.  Therefore, our results for
conserved charges in these backgrounds will reproduce expected results. 
But more generally one could consider cases where higher $\ap$ corrections
contribute. An example is AdS with radius of curvature comparable to the
string scale.  It would be  interesting to apply our approach to such
an example.  

This paper is organized as follows.  In section 2 we show how boundary 
terms arise in the path integral.  This leads to expressions for one-point
functions on the sphere and disk in section 3. A simple example of
the open string with constant gauge field strength is given in section 4. 
In section 5 we review how conserved charges can be derived from the
spacetime action, and this is extended to string theory in section 6. 
Section 7 contains some discussion of open questions.  

\newsec{Boundary terms from the string path integral}

\subsec{Path integral preliminaries}

Although we could presumably be more general, for definiteness
we will consider the general renormalizable bosonic sigma model, following
the conventions of \Polch,
\eqn\aa{\eqalign{S=& {1\over 4\pi\ap}\int_M \! d^2\sigma \, g^{1/2}\left[
\left(g^{ab}G_{\mu\nu}(X)+i \epsilon^{ab}B_{\mu\nu}(X)\right)
\p_a X^\mu \p_bX^\nu + T_c(X)+ \ap R \Phi(X)\right] \cr
 &\quad\quad\quad+ \int_{\p M} \! ds \, \left[i A_\mu(X) 
{dX^\mu \over ds}+ T_o(X)  + 
{k \over 2\pi} \Phi(X) \right].
}}
We will usually be working in $D=25+1$ dimensions, but it is helpful to
keep a general $D$ in most of our formulas.   
It will be important for us to work on a compact worldsheet,
\eqn\aaa{
V_M = \int \! d^2\sigma \, g^{1/2}= ~{\rm finite}.}
  We are interested in 
correlation functions of local operators,
\eqn\ab{\langle {\cal O}_1 \ldots {\cal O}_n \rangle
= \int \! {\cal D}X \,{\cal O}_1 \ldots {\cal O}_n~  e^{-S}.}
The path integral measure has to be treated with some care. We desire
a measure preserving worldsheet diffeomorphism invariance and spacetime
gauge symmetries,
and consistent with the ``ultralocality'' principle 
\refs{\PolchinskiZF,\DHokerTA} of being 
expressible as a pointwise product over the worldsheet.  We can 
formally define the measure in terms of a norm on the tangent space,
\eqn\ac{||\delta X||^2 = \int_M \! d^2\sigma \, g^{1/2} 
G_{\mu\nu}(X) \delta X^\mu \delta X^\nu,}
or as the pointwise product
\eqn\aca{\eqalign{ {\cal D}X = & \prod_\sigma 
\left[-G\left(X(\sigma)\right)\right]^{1/2}
 d^D X(\sigma) \cr
 = &~ e^{ {1 \over 2} \int_M \! d^2\sigma \, g^{1/2}\delta^{(2)}(0)
\ln[ -G \left(X(\sigma)\right)]}\prod_\sigma   d^D X(\sigma).}}
To give meaning to \aca\ we need to define $\delta^{(2)}(0)$ as part of our
regularization procedure.  
%
%
For instance, in heat kernel regularization,
\eqn\acc{\delta^{(2)}(0)~ \Rightarrow~ K_\epsilon(\sigma,\sigma) 
= {1 \over 4\pi \epsilon} + { 1 \over 24 \pi}R + {\cal O}(\epsilon).}
Therefore --- and more generally given ultralocality--- the measure factors
take the form of tachyon and dilaton couplings and so can be absorbed in
\aa\ \refs{\PolchinskiZF,\DHokerTA,\TseytlinTV}.  
Assuming this has been done, in a given regularization scheme 
the tachyon and dilaton might 
therefore transform in an unconventional way under spacetime coordinate
transformations in order to compensate for
the noninvariance of the measure.  This can be rectified by a field
redefinition; that is by adding additional counterterms to \aa.  
 We will assume that the counterterms implicit in \aa\ are such
that spacetime symmetry transformations act in the usual way.  For instance,
this can be made manifest by working with the covariant background field 
expansion.

When the target spacetime is noncompact the path integral \ab\ can be
ill-defined due to the large volume integration.   To isolate this feature we
separate out the integral over the constant mode of $X^\mu$ from the
nonconstant modes.   We therefore
 expand $X^\mu$ in a complete set of modes
\eqn\ad{ X^\mu(\sigma) = x^\mu + \sum_{n\neq 0}
x_n^\mu X_n(\sigma),}
with
\eqn\ae{\int \! d^2\sigma \, g^{1/2} X_n X_m = \delta_{nm}, \quad
\int \! d^2\sigma \, g^{1/2} X_n = 0,
\quad\quad
n,m \neq 0.}
We will sometimes use the notation $  X^\mu = x^\mu+ \tilde{X}^\mu$.
In terms of the mode coefficients the measure is
\eqn\af{ {\cal D}X = (V_M)^{D/2}d^Dx  \prod_{n \neq 0}  d^Dx_n \equiv
(V_M)^{D/2}d^Dx {\cal D}X'.} 
The powers of $V_M$ are due to the different normalization of $x^\mu$ 
compared to 
$x^\mu_n$; these factors are familiar from computations in flat
spacetime.  The normalization of the path integral is in fact fixed
by Weyl invariance and ultralocality of the measure 
\refs{\PolchinskiZF,\DHokerTA}.  

Functional derivatives are defined by $ {\delta \over \delta X^\mu(\sigma)}
X^\nu(\sigma') = g^{-1/2} \delta^{(2)}(\sigma -\sigma')\delta^\nu_\mu$, 
or in terms of modes,
\eqn\ag{{\delta \over \delta X^\mu(\sigma)} = V_M^{-1}{\p \over \p x^\mu}
+ \sum_{n \neq 0} X_n(\sigma) {\p \over \p x_n^\mu} .}

\subsec{Appearance of boundary terms}

In the compact case the classical equations of motion hold inside 
correlation functions since the path integral of a total derivative
vanishes,
\eqn\agb{ \langle {\delta S \over \delta X^\mu(\sigma)} 
{\cal O}_1(\sigma_1) \ldots {\cal O}_n(\sigma_n)\rangle =
- \int \! {\cal D}X\, {\delta \over \delta X^\mu(\sigma)} \left\{
{\cal O}_1(\sigma_1) \ldots {\cal O}_n(\sigma_n) e^{-S} \right\} = 0,}
(modulo contact terms if $\sigma = \sigma_i$).  In the noncompact case
\agb\ can be modified by boundary terms in spacetime.  The point is that the 
factor $e^{-S}$  typically decays exponentially for large 
$|x^\mu_n|$, but this need not
be so for large $|x^\mu|$, for instance as occurs for the case of the
trivial Minkowski vacuum.  Therefore, we should 
only assume that the path integral
of a total derivative with respect to a {\it nonconstant} mode vanishes, and 
so we use \ag\ to write
\eqn\aj{\eqalign{\int \! {\cal D}X \,{\delta \over \delta X^\mu(\sigma)}
\left\{{\cal O}_1 \ldots {\cal O}_n  e^{-S}\right\} & = 
(V_M)^{-1} \int \! {\cal D}X \, {\p \over \p x^\mu} 
\left\{{\cal O}_1 \ldots {\cal O}_n  e^{-S}\right\} \cr
& = (V_M)^{D/2-1} \int \! d^Dx\,{\p \over \p x^\mu}
\langle {\cal O}_1 \ldots {\cal O}_n \rangle ',}}
where $\langle \cdots \rangle '$ denotes the path integral with respect
to the nonconstant modes.


It is  useful to relate the boundary terms to nonholomorphicity
of the energy-momentum tensor. 
The action \aa\ is invariant under the infinitesimal worldsheet diffeomorphism 
\eqn\ak{\delta_\xi g_{ab} = {\nabla_{a}} \xi_{b} + {\nabla_{b}} \xi_{a}, \quad
\delta_\xi X^{\mu} = \xi^{a} \p_a X^{\mu}.}
Using the definition of the energy-momentum tensor,
\eqn\al{T^{ab} = 4\pi {\delta S \over \delta g_{ab}},}
and integrating the variation of the action by parts, we find 
\eqn\am{\nabla^a T_{ab} = 2\pi {\delta S \over \delta X^\mu} \p_b X^\mu.}
\am\ is of course just the statement that the energy-momentum tensor
is conserved when the equations of motion are satisfied.  \am\ holds as
an operator equation once we define  products of fields appropriately.
In terms of  the path integral, the right hand side becomes
\eqn\an{-2\pi \int \! {\cal D}X \, {\delta \over \delta X^\mu(\sigma)}
\left\{ \p_b X^\mu(\sigma) {\cal O}_1 \ldots {\cal O}_n e^{-S}\right\}.}
We are assuming that none of the operators ${\cal O}_i$ is at $\sigma$.
We have also absorbed a delta function contribution into the definition
of the operator ${\delta S \over \delta X^\mu} \p_b X^\mu$.  Indeed, in
free field theory this precisely corresponds to the standard normal
ordering prescription \Polch.  

Now we use \aj\ to write
\eqn\ao{\langle \nabla^a T_{ab}(\sigma) {\cal O}_1 \ldots {\cal O}_n \rangle
= -2\pi (V_M)^{D/2-1} \int \! d^Dx \,  {\p \over \p x^\mu}\langle 
\p_b X^\mu(\sigma) {\cal O}_1 \ldots {\cal O}_n \rangle '.}
Despite appearances, \ao\ is invariant under spacetime diffeomorphisms since,
 according to our measure conventions, 
$\langle \p_b X^\mu(\sigma) {\cal O}_1 \ldots 
{\cal O}_n \rangle '$ transforms like $(-G)^{1/2}$ times a spacetime vector.

One is used to saying that the left hand side of \ao\ should vanish by
worldsheet diffeomorphism invariance. But as we have shown, this 
conclusion only follows if the classical equations of motion hold
inside correlators, and this can be violated by boundary terms. \ao\
is consistent with worldsheet diffeomorphism invariance.

\newsec{One point functions on the sphere and disk}

\subsec{The sphere}

We now restrict to Weyl invariant theories of the form \aa, including  also the
Fadeev-Popov determinant to cancel the matter central charge.  
We work in conformal gauge
\eqn\ba{ds^2 = e^{2\omega(z,\zb)}dz d\zb.}
Now consider a local operator of scaling dimension $(h,\tilde{h})$ obeying
the standard OPE
\eqn\bb{ T(z',\bar{z}') {\cal O}(z,\bar{z}) \sim \ldots + 
{h {\cal O}(z,\bar{z})
\over (z' -z)^2} + {\p  {\cal O}(z,\bar{z}) \over z'-z},}
and similarly for $\tilde{T}(z',\zb')$.   It is important to emphasize that
\bb\ should hold on a {\it curved} worldsheet.  On a curved worldsheet 
operators of different engineering dimension can mix via appearance of
factors of $R_{z\zb}$, so an operator of definite scaling dimension
on a flat worldsheet need not have definite scaling dimension on a 
curved worldsheet.   

Let $C$ be a small contour circling $z$.  Using the OPE we have for the
one point function with $h\neq 0$:
\eqn\bc{\langle {\cal O}(z,\zb) \rangle_{S_2} = {1 \over h} 
\oint_C {dz' \over 2\pi i} (z'-z)\langle T(z',\bar{z}') {\cal O}(z,\zb)
\rangle_{S_2}.}
We would now like to deform the contour, eventually contracting it to zero
by ``sliding it off the opposite pole of the sphere''.
In a compact CFT the stress tensor is holomorphic, and therefore the 
correlator is independent of $C$ provided that no other 
operators are encountered.
This is the standard logic by which one concludes that all one point functions
of operators with $h\neq 0$ vanish on the sphere.  But in a noncompact
CFT the stress tensor can be non-holomorphic as in \ao, 
and so we will pick up a
contribution from deforming the contour.  
In particular, we use the divergence theorem
\eqn\bd{\int_R \! d^2z\, (\p v^z + \bar{\p}v^{\zb}) 
= i \oint_{\p R}(v^z d\zb - v^{\zb}dz),}
to write
\eqn\be{\langle {\cal O}(z,\zb) \rangle_{S_2} =
{1 \over 2\pi h} \int_{S_2} \! d^2z'\, (z'-z)
\langle\bar{\p} T(z',\bar{z}'){\cal O}(z,\zb)\rangle_{S_2}.}
Then using \ao\ we arrive at
\eqn\bg{\langle {\cal O}(z,\zb) \rangle_{S_2} = 
-{1 \over 2 h} (V_M)^{D/2-1}\int \! d^Dx \,  \p_\mu
\left\{ \int \! d^2z'\, (z'-z)e^{2\omega(z',\bar{z}')}\langle
\p X^\mu(z',\bar{z}'){\cal O}(z,\zb)\rangle'_{S_2}\right\}.}
\bg\ gives our desired result: it expresses a one point function in the CFT as 
a boundary term in spacetime.

\subsec{The disk}

Now consider a worldsheet of disk topology.  A conformal field theory
on the disk has an energy-momentum tensor obeying the boundary condition
\eqn\ca{n^a t^b T_{ab}|_{\p D_2} =0,}
where $n^a$ and $t^a$ are normal and tangent to the  boundary.  The conformal
symmetry is generated by a single copy of the Virasoro algebra.   To derive
the form of the Virasoro generators it is convenient to start with a
representation of the disk as the upper half $w$ plane, with a metric chosen
such that the worldsheet volume is finite.  We again work in conformal
gauge \ba.  The boundary conditions are then
\eqn\cb{ T(w) = \tilde{T}(\bar{w}), \quad w=\bar{w}.}
Let $C$ be any contour in the $w$ plane with endpoints on the real axis.
Then, assuming for the moment holomorphicity of $T$,  the following charges
are independent of the contour $C$ provided that no other operators are
encountered,
\eqn\cc{L_n = \int_C \left[ {dw' \over 2\pi i} (w')^{n+1}T(w')
-{d\bar{w}' \over 2\pi i} (\bar{w}')^{n+1}\tilde{T}(\bar{w}')\right].}

For calculational purposes it can be convenient to use a flat worldsheet
metric.  We also want a finite coordinate range, so we rewrite the above 
charges after transforming to the 
disk $|z'| \leq 1$.  Let $z$ be a point on the boundary, $|z|=1$, and let
the map from the $w'$-plane to the $z'$-plane be
\eqn\cd{ z' =  \left({i-w' \over i+w'}\right) z.}
The Virasoro generators then take the form
\eqn\ce{L_n = \int_C \left[{dz' \over 4\pi z}(z+z')^2
 \left(i{z-z' \over z+z'}\right)^{n+1}T(z')
+ {d\bar{z}' \over 4\pi \zb}(\zb+\zb')^2 
\left(-i{\zb-\zb' \over \zb+\zb'}\right)^{n+1}\tilde{T}(\zb')
\right].}
$C$ is now any contour with endpoints on the boundary of the disk.  

In CFTs with noncompact target spaces, the charges $L_n$ need not be
independent of $C$ since the energy-momentum tensor need not be 
holomorphic.   As we did for the sphere, we use this to give a formula
for one point functions of boundary operators in terms of surface 
integrals in the target space.   So consider a local boundary operator
${\cal O}(z,\zb)$ with scaling dimension $h \neq 0$.  If we let $C$ be a tiny
semi-circular contour around $z$ then we can use the OPE to write
\eqn\cf{L_0 {\cal O}(z,\zb) = h {\cal O}(z,\zb).}
Using the divergence theorem gives
\eqn\cg{\eqalign{ \langle {\cal O}(z,\zb) \rangle_{D_2} =
{1 \over 2\pi h} \int_{D_2}\!d^2z'\,& \left\{
\left({z'+z\over 2 z}\right)
 (z'-z)\langle \bar{\p} T(z',\bar{z}'){\cal O}(z,\zb) \rangle_{D_2} \right.\cr
&+\left. \left({\zb'+\zb\over 2 \zb}\right) (\zb'-\zb)  
\langle \p\tilde{T}(z',\zb')
{\cal O}(z,\zb)\rangle_{D_2} \right\}.}}
Then  \ao\ gives the final result
\eqn\ch{\eqalign{ \langle {\cal O}(z,\zb) \rangle_{D_2} =&
-{1 \over 2 h} (V_M)^{D/2-1}\int \! d^Dx \, \p_\mu
\left\{ \int_{D_2} \! d^2z'\,e^{2\omega(z',\bar{z}')}\right. \cr
&\left.\left({z'+z\over 2 z}\right)
 (z'-z)\langle  \p X^\mu(z',\bar{z}'){\cal O}(z,\zb) \rangle'_{D_2}+ 
\left({\zb'+\zb\over 2 \zb}\right)
 (\zb'-\zb)\langle  \bar{\p} X^\mu(z',\bar{z}'){\cal O}(z,\zb) \rangle'_{D_2}
\right\}.
}}

\newsec{Example: open string with constant field strength}

 The simplest illustration of our result is for the open string with 
constant field strength.   We work on the unit disk $|z| \leq 1$
with flat metric $ds^2 = dzd\zb$.   The sigma model action is
\eqn\da{S= {1\over 2\pi\ap}\int_{D_2}\! d^2z \,
\eta_{\mu\nu}\p X^\mu \bar{\p}X^\nu +{i \over 2}\oint_{\p D_2} \! d\theta 
F_{\mu\nu}X^\mu {\p X^\nu \over \p \theta}.}
The propagator 
\eqn\db{G^{\mu\nu}(z',\zb';z,\zb) = {\langle
X^\mu(z',\zb')X^\nu(z,\zb) \rangle'_{D_2} \over 
\langle 1 \rangle'_{D_2}} - x^\mu x^\nu}
obeys
\eqn\dc{\p \bar{\p} G^{\mu\nu} = -\pi \ap  \left[
\delta^{(2)}(z' -z) - {1 \over 2\pi} \right] \eta^{\mu\nu},}
with boundary condition
\eqn\dd{\left[{\p \over \p r'}G^{\mu\nu} + 2\pi i \ap F^\mu_{~\,\alpha} 
{\p \over \p \theta'} G^{\alpha \nu}\right]_{r' =1} = 0.
}
We wrote $z = r e^{i\theta}$. 
\lref\AbouelsaoodGD{
A.~Abouelsaood, C.~G.~Callan, C.~R.~Nappi and S.~A.~Yost,
Nucl.\ Phys.\ B {\bf 280}, 599 (1987).
}
The solution is \AbouelsaoodGD,
\eqn\de{G^\mu_{~\nu}(z',\zb';z,\zb) =
{\ap \over 2} \left[ - \ln|z'-z|^2 - { 1 + {\cal F} \over 1 - {\cal F}}
\ln(1-z' \zb) - { 1 - {\cal F} \over 1 + {\cal F}}
\ln(1-\zb' z) + z'\zb' + z\zb +c \right]^\mu_{~\nu}}
where ${\cal F} = 2\pi \ap F$. 
The constant $c$ is fixed by requiring $\int \! d^2z \, G^{\mu\nu}=0$,
but we will not need its value.  From \de\ it is straightforward to see
that $T = -{1  \over \ap} : \p X^\mu \p X_\mu:$ is not holomorphic inside a 
general correlation function.

Now we use our formula \ch\ to compute the following one-point function
\eqn\df{F_{\mu\nu}\langle X^\mu  {\p X^\nu \over \p \theta} \rangle_{D_2}.}
This is clearly a ``nonnormalizable operator'', since it corresponds
to a gauge field which becomes arbitrarily large in spacetime.  
Of course, it is easy to compute \df\  directly from \de,
but we use \ch\ for illustration.  
Acting with $\p_\mu$ we find
\eqn\dg{\eqalign{F_{\mu\nu}\langle X^\mu  {\p X^\nu \over \p \theta} 
\rangle_{D_2}=
-&{1 \over  2}(V_M)^{D/2-1}\langle 1 \rangle'_{D_2} F_{\mu\nu}
\int \! d^Dx \, \int_{D_2} \! d^2z' \cr
&\left\{
\left({z'+z\over 2 z}\right)
 (z'-z)  {\p \over \p z'}{\p \over \p \theta}G^{\mu\nu}+
\left({\zb'+\zb\over 2 \zb}\right)
 (\zb'-\zb){\p \over \p \zb'}{\p \over \p \theta}G^{\mu\nu} \right\}.}}
Using \de\ and performing the $z'$ integral gives
\eqn\dh{F_{\mu\nu}\langle X^\mu  {\p X^\nu \over \p \theta} 
\rangle_{D_2}={i \over 4 \pi} \eta_{\mu\nu}
\left[{(2\pi \ap F)^2 \over 1 - (2\pi \ap F)^2}
\right]^{\mu\nu} \langle 1 \rangle_{D_2}.}
We have used 
\eqn\di{ \langle 1 \rangle_{D_2} = 
(V_M)^{D/2}\int \! d^Dx\, \langle 1 \rangle'_{D_2},}
(see \af).

\ From \da\ it follows that
\eqn\dj{F_{\mu\nu}\langle X^\mu  {\p X^\nu \over \p \theta} 
\rangle_{D_2} =  {i \over 2\pi} F_{\mu\nu}
{\delta \over \delta F_{\mu\nu}} \langle 1 \rangle_{D_2}.}
Integrating then gives
\eqn\dk{\langle 1 \rangle_{D_2} = N \int \! d^Dx\,
\sqrt{-\det[\eta_{\mu\nu}+2\pi\ap F_{\mu\nu}]}.}
The Born-Infeld action \dk\ is  indeed the expected result
for the partition function in the presence of a constant field
strength \TseytlinDJ, so our result \dh\ for the one-point function is
correct.  

This example is ``trivial'' in the sense that \da\ is a free theory and
so all correlators are easily computed without use of \ch.  More generally,
we will have an interacting CFT which simplifies at infinity in spacetime,
in which case \ch\ is needed.  We  exploit this below in our derivation of
conserved charges. 

\newsec{Spacetime action and conserved charges}

Conserved charges associated with gauge symmetries appear as surface
integrals at spatial infinity.  This is most easily seen by considering
the variation of the spacetime action with respect to gauge transformations
supported at infinity \refs{\BrownBR,\WaldWA}.

  Let $A_p$ and $\phi_q$ denote
some collection of gauge and matter fields, with all spacetime indices
suppressed.   Consider a gauge invariant spacetime action for these fields,
\eqn\ea{S = \int_V \! d^Dx \, { \cal L}_{{\rm bulk}}(A_p,\phi_q) + 
\int_{\p V} \! d^{D-1}x \, {\cal L}_{{\rm bndy}}(A_p,\phi_q).}
Both ${ \cal L}_{{\rm bulk}}$ and ${\cal L}_{{\rm bndy}}$  are allowed to 
depend on first and higher derivatives acting on the fields.   In general,
the boundary term is required for two reasons.  First, the Euler-Lagrange
equations should imply stationarity of the action with respect to variations
that vanish at the boundary; if ${ \cal L}_{{\rm bulk}}$ contains second
or higher derivatives then an appropriate ${\cal L}_{{\rm bndy}}$ will
be needed to cancel terms arising from integration by parts.  A familiar
example is the Gibbons-Hawking boundary term \GibbonsUE\ which is
 added to the Einstein-Hilbert
action.   Second, the action might diverge as the boundary is taken to infinity
within the class of field configurations that one wishes to include in the 
theory.  If so, one can try to add an additional boundary term to cancel this
divergence, though this boundary term should depend only on the boundary 
values of the  fields and not their normal derivatives, otherwise the
Euler-Lagrange equations will no longer follow.  Boundary terms of this
sort arise naturally in anti-de Sitter spacetime 
\refs{\HenningsonGX,\BalasubramanianRE, \EmparanPM,\KrausDI}. 

Given an acceptable spacetime action, consider a general field variation
about some configuration satisfying the Euler-Lagrange equations. 
The bulk term vanishes by assumption, leaving the boundary term
\eqn\eb{\delta S = \int_{\p V} \! d^{D-1}S \,\left\{
\pi_{A_p}\delta A_p + \pi_{\phi_q}\delta \phi_q\right\}.}
If $\p V$ is a constant time hypersurface then $\pi_{A_p}$ and $\pi_{\phi_q}$
define the usual canonical momenta; more generally they are functionals
of the fields and their normal and tangential derivatives.  

To define conserved charges let $\p V$ be a timelike surface at spatial
infinity, and  consider a gauge transformation parameterized
by $\xi$.   In general, both the gauge and matter fields will contribute
a nonzero variation.  However, one usually 
imposes asymptotic conditions on the matter fields such that
\eqn\ec{ \int_{\p V} \! d^{D-1}S \,\pi_{\phi_q}\delta_\xi \phi_q=0.}
Assuming that this is the case, then the  gauge invariance of the action,
$\delta_\xi S =0$, implies
\eqn\ed{ \int_{\p V} \! d^{D-1}S \,  \pi_{A_p}\delta_\xi A_p = 0.}
Now take $\xi$ to depend only on time with respect to some asymptotic
timelike Killing vector, and to be tangent to $\p V$ (when $\xi$ has 
spacetime indices).
  Integrating by parts so that $\xi$ appears
without derivatives we will arrive at an expression of the form 
\eqn\ee{\int \! dt \, \xi(t) {d Q \over dt} = 0.}
Since $\xi(t)$ is arbitrary we obtain the conserved charge $Q$.  

As a rather elementary example consider an abelian gauge field 
minimally coupled to a complex scalar field,
\eqn\ef{S = \int \! d^Dx \, \left\{ -{1 \over 4}F^{\mu\nu}F_{\mu\nu}
+ {1 \over 2}(D^\mu \phi)^* D_\mu \phi- V(\phi^* \phi)\right\}.}
When the equations of motion are satisfied the variation of the action
is
\eqn\eg{\delta S = \int \! dt d^{D-2}S_i \left\{-F^{i\mu}\delta A_\mu 
+{1\over 2} (D^i \phi)^* 
\delta \phi + {1\over 2}D^i \phi \delta \phi^*\right\}.}
Consider a gauge transformation, $\delta_\xi A_\mu = \p_\mu \xi$, 
$\delta_\xi \phi = i \xi \phi$, with $\xi = \xi(t)$.  Then, assuming that
the matter current $J^i = i[\phi^* D^i \phi - \phi (D^i \phi)^*]/2$ falls
off faster than $1/r^{D-2}$, we obtain the usual expression for electric
charge
\eqn\eh{\delta S = -\int \! dt d^{D-2}S_i\, F^{i0}{d\xi \over dt}
\quad \Rightarrow \quad Q ~\propto ~\int \!d^{D-2}S^i\, F^{i0}.}

As our main example consider the low energy action for the bosonic
string
\eqn\elA{
S = S_V + S_{\partial V} + S_{ct}
}
where 
\eqn\el{S_V= {1 \over 2 \kappa_0^2} \int \! d^{D} x (-G)^{1/2} e^{-2 \Phi}
\left\{ R - {1 \over 12} H_{\mu\nu\lambda}H^{\mu\nu\lambda}
+4 \p_\mu \Phi \p^\mu \Phi \right\}}
is the standard bulk action, 
and we have to add the Gibbons-Hawking surface term 
\eqn\ela{
S_{\partial V}= - {1 \over \kappa_0^2} \int_{\partial V} 
\! d^{D-1} x (-\gamma)^{1/2} e^{-2 \Phi} \Theta .}
$S_{ct}$ is required for finiteness of the action, but its precise form 
will not be needed; see 
\refs{\HenningsonGX,\BalasubramanianRE, \EmparanPM,\KrausDI} for more details.
Here 
\eqn\elb{
\gamma_{\mu\nu} = G_{\mu\nu} - n_\mu n_\nu 
}
is the induced metric on the (assumed to be timelike) boundary $\partial V$, 
\lref\Wald{
R. Wald, 
``General relativity,''
The University of Chicago Press (1984), Chapters 9 and 10 
(particularly Section 10.2).}
and the extrinsic curvature tensor and scalar 
\eqn\elc{
\Theta_{\alpha\beta} = - \gamma_{\alpha\mu} \nabla^\mu n_\beta
,\quad
\Theta = - \gamma^{\mu\nu} \nabla_\mu n_\nu
}
are defined in the standard way, see \BrownBR, \Wald. 
Indices are raised and lowered 
with the original metric $G_{\mu\nu}$, and 
$n^\mu$ is the outward unit normal vector to $\partial V$ 
($n^\mu$ is spacelike).
The extrinsic curvature satisfies 
$\Theta_{\alpha\beta} n^\beta = \Theta_{\beta\alpha} n^\beta = 0$, 
from which follows 
$\Theta_{\alpha\beta} = \Theta_{\beta\alpha}$ 
(we also assume $n^\alpha \nabla_\alpha n_\beta = 0$;
for further details see \Wald).
Then the definition \elc~ agrees with 
the manifestly symmetric expression
\eqn\elca{
\Theta^{\alpha\beta} = - 
{1\over 2} (\nabla^\alpha n^\beta+\nabla^\beta n^\alpha)
}
used in \BalasubramanianRE~ and elsewhere.
Finally, the volume elements on $V$ and $\partial V$ 
\eqn\elcc{\eqalign{
(d^{D} x) (-G)^{1/2} &= \epsilon_{\mu_1 ... \mu_D} \, 
dx^{\mu_1} \wedge ... \wedge dx^{\mu_D}
, \cr
(d^{D-1} x) (-\gamma)^{1/2} &= \tilde\epsilon_{\mu_2 ... \mu_D} \, 
dx^{\mu_2} \wedge ... \wedge dx^{\mu_D}
, \cr
{1\over D} \epsilon_{\mu_1 \mu_2 ... \mu_D} &= 
n_{[\mu_1} \tilde\epsilon_{\mu_2 ... \mu_D ]}
}}
are related through Gauss's law
\eqn\elccc{\eqalign{
\int_V d^{D} x (-G)^{1/2} \nabla_\mu v^\mu & = 
\int_{\partial V} d^{D-1} x (-\gamma)^{1/2} n_\mu v^\mu \cr
& \equiv \int_{\partial V} d^{D-1}S_\mu v^\mu.}
}

Setting $S_{ct} = 0$ for the moment, the variation of the action \elA~ is a 
bulk term which 
vanishes when the equations of motion 
\eqn\eld{\eqalign{
0 &=
R - {1\over 12} H^2 + 4 \nabla^2 \Phi - 4 (\nabla \Phi)^2
\cr
0 &= 
{1\over 2} \nabla^\alpha H_{\mu\nu\alpha} - 
\nabla^\alpha \Phi H_{\mu\nu\alpha} 
\cr
0 &= R_{\mu\nu} - {1\over4} H_{\alpha\beta\mu} H^{\alpha\beta}{}_{\nu} 
+ 2 \nabla_\mu\nabla_\nu\Phi
}}
are satisfied, plus the boundary term
\eqn\ele{\eqalign{
\delta S = {1 \over 2\kappa_0^2} \int_{\partial V} 
\! d^{D-1} x (-\gamma)^{1/2} e^{-2 \Phi} 
\Bigg\{\Bigg.&
4 \delta \Phi (\Theta + 2 n^\mu \nabla_\mu \Phi )
-{1\over2} \delta B_{\alpha\beta} n_\gamma H^{\alpha\beta\gamma}
\cr&
- \delta G_{\alpha\beta} 
\left[ 
\gamma^{\alpha\beta} (\Theta + 2 n^\mu \nabla_\mu \Phi )
- \Theta^{\alpha\beta}) 
\right]
\Bigg.\Bigg\}.
}}

There are conserved gauge charges associated with diffeomorphisms
as well as anti-symmetric tensor gauge transformations. 
Consider first $\delta \Phi = \delta G_{\mu\nu} = 0$, and the gauge
transformation
\eqn\elf{
\delta_\Lambda B_{\mu\nu} =
 \nabla_\mu \Lambda_\nu - \nabla_\nu \Lambda_\mu 
= \partial_\mu \Lambda_\nu - \partial_\nu \Lambda_\mu.
}
 Then \ele~ gives 
\eqn\elg{
\delta S = {1 \over 2\kappa_0^2} \int_{\partial V} 
\! d^{D-1}S_\gamma\,  e^{-2 \Phi} 
\Bigg\{\Bigg.
- (\nabla_\alpha \Lambda_\beta) H^{\alpha\beta\gamma}
\Bigg.\Bigg\}
}
We take the metric to approach the Minkowski metric asymptotically, so that
it makes sense to talk about time at spatial infinity. 
Assuming this 
is the case, we can take our gauge parameter 
to be a function of only $t$ at spatial infinity, 
so demanding that $\delta S=0$ 
gives rise to the conserved charges 
\eqn\elgc{Q_j ~ \propto ~ \int \! d^{D-2}S^i \, e^{-2 \Phi}H_{0ij}.}
%
Here and in the following,  $D-2$ dimensional integrals are evaluated
on a constant $t$ slice of $\p V$. 

Alternatively, consider a diffeomorphism.   
The fields transform as 
\eqn\elh{\eqalign{
\delta_\xi G_{\mu\nu} &= \nabla_\mu \xi_\nu + \nabla_\nu \xi_\mu
\cr
\delta_\xi \Phi &= \xi^\alpha \nabla_\alpha \Phi
\cr
\delta_\xi B_{\mu\nu} &= \xi^\alpha \nabla_\alpha B_{\mu\nu} + 
\nabla_\mu \xi^\alpha B_{\alpha\nu} + 
\nabla_\nu \xi^\alpha B_{\mu\alpha} 
\cr&
=\xi^\alpha H_{\alpha\mu\nu} + 
\left[
\nabla_\mu (\xi^\alpha B_{\alpha\nu}) - 
\nabla_\nu (\xi^\alpha B_{\alpha\mu}) 
\right].
}}
The term in square brackets in the last line of \elh\ is 
of the form \elf.  It is a gauge variation of the $B_{\mu\nu}$ field 
and so can be dropped. 
Now \ele~ 
becomes 
\eqn\eli{\eqalign{
\delta S &= -{1 \over 2\kappa_0^2} \int_{\partial V} 
\! d^{D-1} x (-\gamma)^{1/2} e^{-2 \Phi} 
\Big\{\Big.
2 \nabla_\alpha \xi_\beta 
\left[ 
\gamma^{\alpha\beta} (\Theta + 2 n^\mu \nabla_\mu \Phi )
- \Theta^{\alpha\beta} 
\right]
\cr&\quad\quad\quad\quad\quad\quad\quad\quad\quad
- \xi_\beta
\left[
4 \nabla^\beta \Phi (\Theta + 2 n^\mu \nabla_\mu \Phi )
-{1\over2} H^{\beta\mu\nu} n^\gamma H_{\mu\nu\gamma}
\right]
\Big.\Big\}.
}}
If $\nabla_\mu \Phi$ and $H_{\mu\nu\lambda}$ 
vanish sufficiently rapidly at infinity they will not
contribute to the surface integrals (this is the condition that there
is no matter flux out through spatial infinity), 
and the second line in \eli\ vanishes. 
A conserved energy-momentum vector is obtained by taking 
$\xi_\mu = \xi_\mu(t)$
at the boundary,
\eqn\elia{\eqalign{
P^\beta &~\propto ~ 
- \int 
\! d^{D-2} x\, (-\gamma)^{1/2} e^{-2 \Phi} 
\left[ 
\gamma^{0\beta} (\Theta + 2 n^\mu \nabla_\mu \Phi )
- \Theta^{0\beta} 
\right].
}}
A conserved angular momentum can be obtained in a similar way  
by allowing for spatial dependence
in $\xi_\mu$  at infinity.

Actually, \elia\ diverges even for empty Minkowski space as the
boundary is taken to infinity, which  is why one needs to include the
counterterm action
$S_{ct}$ in \elA.  The form of $S_{ct}$ depends on the asymptotic boundary 
conditions that have been chosen; examples for asymptotically flat
and asymptotically AdS spacetimes can be found in 
\refs{\HenningsonGX,\BalasubramanianRE,\EmparanPM,\KrausDI}.  In the
asymptotically flat case  the effect is to 
simply subtract from \elia\ the terms which are nonvanishing in 
empty Minkowski space.

It is instructive to compare expressions 
written in the string frame, 
such as \eli, 
with their counterparts in the Einstein frame. 
Metrics in the string and Einstein frames are related as 
\eqn\ema{
G_{\mu\nu}^E = e^{2 \omega} G_{\mu\nu}
,\quad\quad
\omega \equiv - {2 \tilde \Phi \over D-2}
,\quad\quad
\tilde \Phi \equiv \Phi - \Phi_0 
,\quad\quad
\kappa \equiv \kappa_0 e^{\Phi_0}. 
}
This filters through to the definitions of the unit 
normal vector to $\partial V$ and the connection
\eqn\emb{\eqalign{
n_\mu^E &= e^{\omega} n_\mu
\cr
\Gamma^E{}^\alpha{}_{\mu\nu} &= \Gamma^\alpha{}_{\mu\nu} 
+\left( 
\delta^\alpha_\mu \, \partial_\nu \omega + 
\delta^\alpha_\nu \, \partial_\mu \omega - 
G_{\mu\nu} \, G^{\alpha\beta} \, \partial_\beta\omega 
\right).
}}
\ From this, we see that the extrinsic curvatures in the two frames 
are related as 
\eqn\emd{\eqalign{
\Theta_{\mu\nu}^E &= e^{\omega} 
\left( 
\Theta_{\mu\nu}
- 
\gamma_{\mu\nu} \, n^\beta \, \nabla_\beta\omega
\right)
, \cr
\Theta^E &= e^{-\omega} 
\left( 
\Theta
- 
(D-1) \, n^\beta \, \nabla_\beta\omega
\right)
}}
and indices are raised and lowered with the metric 
appropriate for a given frame. 
The  expression for the 
gravitational stress-energy 
in Einstein frame converts to string frame as 
\eqn\eme{
\Theta_E \gamma_E^{\mu\nu} - \Theta_E^{\mu\nu}
= 
e^{6 \tilde \Phi \over D-2} 
\left(
\Theta \gamma^{\mu\nu} - \Theta^{\mu\nu}
+ 
2 \gamma^{\mu\nu} \, n^\beta \, \nabla_\beta\Phi
\right).
}
Finally, diffeomorphism parameters in the two 
frames
are related as  
\eqn\emg{\eqalign{
\xi_\mu^E &= e^{2 \omega} \xi_\mu 
\cr
\delta_\xi G^E_{\mu\nu} &=e^{2 \omega} 
\left(
\delta_\xi G_{\mu\nu}
+ 2 G_{\mu\nu} \xi^\beta \, \partial_\beta\omega
\right).}}
Hence the first line of \eli\
reads 
\eqn\emh{\eqalign{
\delta S 
&= 
- {1 \over 2\kappa_0^2} \int_{\partial V} 
\! d^{25} x (-\gamma)^{1/2} e^{-2 \Phi} 
\Big\{\Big.
2 \nabla_\alpha \xi_\beta 
\left[ 
\gamma^{\alpha\beta} (\Theta + 2 n^\mu \nabla_\mu \Phi )
- \Theta^{\alpha\beta}) 
\right]
\Big.\Big\}
\cr
&
= 
- {1 \over 2\kappa^2} \int_{\partial V} 
\! d^{25} x (-\gamma_E)^{1/2} 
\Big\{\Big.
2 \nabla^E_\alpha \xi^E_\beta 
\left[ 
\gamma_E^{\alpha\beta} \Theta_E - \Theta_E^{\alpha\beta} 
\right]
+ 4 \, \xi_E^\mu \nabla_\mu \tilde \Phi \, \Theta_E 
\Big.\Big\}.
}}
The last term 
is again a flux, and does not contribute 
to the conserved charges provided $\nabla_\mu \Phi$ 
falls off sufficiently rapidly at spatial infinity.  From the 
term proportional to $\nabla^E_\alpha \xi^E_\beta$
we can read off the  energy-momentum vector 
in the Einstein frame, and this is known to agree with the 
standard ADM definition.  Indeed, the definition of energy-momentum
is unique if one demands that $P^\mu$ is conserved, transforms 
like a Lorentz vector under asymptotic Lorentz transformations, and
is additive for distant subsystems \weinberg.

\newsec{Conserved charges in string theory}

We would like to repeat the analysis of the previous section in the
context of the string path integral.   The basic idea is to examine
the behavior of the partition function under spacetime gauge transformations.

\subsec{Open string}

We first  consider the case of the open string with a nontrivial
gauge field
\eqn\fa{S= {1\over 2\pi\ap}\int_{D_2}\! d^2z \,
\eta_{\mu\nu}\p X^\mu \bar{\p}X^\nu +i \oint_{\p D_2} \! d\theta\, 
A_\mu(X){\p X^\mu \over \p \theta}.}
We are working on the unit disk 
with flat metric and boundary coordinate $\theta$.  Spacetime gauge
invariance corresponds to the fact that $\delta_\xi A_\mu = \p_\mu \xi$ 
simply adds a total derivative to the worldsheet Lagrangian, which then
integrates to zero provided one chooses a suitable regularization
scheme.  

Now, consider a gauge field variation such that ${\cal O} = \delta A_\mu(X)
{\p X^\mu \over \p \theta}$ is a dimension $h=1$ boundary operator;
 {\it i.e.} such that $\delta A_\mu$ satisfies the linearized spacetime
equations of motion expanded around $A_\mu$.  \ch\ gives
us  a formula for the one point function of ${\cal O}.$  Since the
one point function is expressed as a boundary term at infinity, what matters
is the behavior of $A_\mu$ and $\delta A_\mu$ for large values of the 
constant modes $x^i$.   Therefore we expand in powers of nonconstant modes,
writing $X^\mu = x^\mu + \tilde{X}^\mu$, where $\tilde{X}^\mu$ are the 
nonconstant modes as in \ad.   
At leading order we can write
\eqn\fbb{\delta A_\mu(X){\p X^\mu \over \p \theta}  = 
\delta A_\mu(x){\p \tilde{X}^\mu \over \p \theta}+ \ldots~.}
Higher order terms in the expansion will be seen to give a vanishing
contribution provided we assume that a derivative expansion is valid at
spatial infinity (that is, that $\p_\mu \sim 1/r$).
Applying \ch\ we have
\eqn\fbc{\eqalign{\big\langle 
& \delta A_\mu(X) {\p X^\mu \over \p \theta}(z,\zb)
\big\rangle_{D_2} 
 = -{1 \over 2} (V_M)^{D/2-1}\int \!d^{D-1}S^i \, \delta A_\mu(x)
\int \! d^2 z'\, \cr 
& \left\{ \left({z'+z\over 2 z}\right)
 (z'-z)\big\langle   \p X^i(z',\bar{z}'){\p \Xt^\mu \over \p \theta}(z,\zb)
 \big\rangle'_{D_2}+ 
\left({\zb'+\zb\over 2 \zb}\right)
 (\zb'-\zb)\big\langle  \bar{\p} X^i(z',\bar{z}')
{\p \Xt^\mu \over \p \theta}(z,\zb) \big\rangle'_{D_2}
\right\}. }}
Now we take $\delta A_\mu = \p_\mu \xi(X^0)$.  The left hand side of \fbc\
vanishes since it is the $\theta$ derivative of a $\theta$ independent
quantity.  Upon integrating the right hand side by parts, we get an 
equation stating that the following charge is conserved
\eqn\fbd{\eqalign{Q ~ = ~ \int  \!d^{D-2}S^i \,\int \! d^2 z'\, 
&  \left\{ \left({z'+z\over 2 z}\right)
 (z'-z)\big\langle   \p X^i(z',\bar{z}'){\p \Xt^0 \over \p \theta}(z,\zb)
 \big\rangle'_{D_2}  \right.   \cr
&\quad +\left. \left({\zb'+\zb\over 2 \zb}\right)
 (\zb'-\zb)\big\langle  \bar{\p} X^i(z',\bar{z}')
{\p \Xt^0 \over \p \theta}(z,\zb) \big\rangle'_{D_2}
\right\}. }}

\fbd\ gives our result for the conserved electric charge in string
theory.  To check it we now
compute $Q$ in a background such that all gauge invariant combinations
of fields  die
off at spatial infinity; {\it i.e.} a localized charge/current distribution. 
At infinity we can therefore expand $A_\mu$ as 
\eqn\fbe{\oint_{\p D_2} \! d\theta\, A_\mu(X){\p X^\mu \over \p \theta}
= {1 \over 2} F_{\mu\nu}(x)\oint_{\p D_2} \! d\theta\, \tilde{X}^\mu
{\p \tilde{X}^\nu \over \p \theta}+ \ldots, 
}
Discarded terms will not contribute to $Q$.  
Since $F_{\mu\nu}(x)$ goes to zero  at infinity,
at the boundary it makes sense to expand in powers of $F_{\mu\nu}(x)$,
hence we need only compute correlation functions in the free CFT on the disk.
The contribution to $Q$ at zeroth order in 
$F_{\mu\nu}$ is easily seen to vanish,
so the leading nonzero contribution is
\eqn\fbf{\eqalign{Q ~ = & ~{1 \over 2} \int  \!d^{D-2}S^i \,F_{\alpha\beta}(x)
\int \! d^2 z'\int \! d\theta'' \, \cr
&\quad \left\{ \left({z'+z\over 2 z}\right)
 (z'-z)\big\langle   \p X^i(z',\bar{z}'){\p \Xt^0 \over \p \theta}(z,\zb)
 \tilde{X}^\alpha(z'',\zb'')
{\p X^\beta \over \p \theta}(z'',\zb'')\big\rangle'_{0,D_2}  \right.  \cr  
 & \quad +\left. \left({\zb'+\zb\over 2 \zb}\right)
 (\zb'-\zb)\big\langle  \bar{\p} X^i(z',\bar{z}')
{\p \Xt^0 \over \p \theta}(z,\zb) \tilde{X}^\alpha(z'',\zb'')
{\p X^\beta \over \p \theta}(z'',\zb'')\big\rangle'_{0,D_2}
\right\} \cr
& = 4\pi^2 (\ap)^2  \langle 1 \rangle'_{D_2}\int \! d^{D-2}S^i\,
F^{0i}(x),}}
where $\langle \ldots \rangle_0$ indicates that 
expectation values are with respect to the first term of \fa.
We therefore find a conserved charge
of the same form as in \eh:
\eqn\fd{ Q ~\propto ~
\int \!d^{D-2}S^i\, F^{0i}.}
For this to be finite  $F^{r0}$ must fall off as $1/r^{D-2}$,
and so all higher powers in the expansion will give vanishing contributions.
Not surprisingly, the conserved charge \fd\ derived from string theory
has exactly the same form as in field theory.   We should emphasize that
to reach this conclusion 
we assumed that all gauge invariant fields and derivatives  die
off asymptotically; if these conditions are relaxed one would expect
to find corrections to low energy field theory.

\subsec{Closed string}

In the closed string the relevant symmetries are anti-symmetric
tensor gauge transformations and spacetime coordinate transformations.
We consider asymptotically flat backgrounds in which $H_{\mu\nu\lambda}$,
$T_c$, and $\nabla_\mu \Phi$ go to zero asymptotically.
Any asymptotically
flat spacetime has -- by definition -- a metric that can be brought 
to the form \mtw\
\eqn\ha{\eqalign{ds^2 & = - \left( 1 - {\mu \over r^{D-3}} 
+ {\rm O}\left({1 \over
r^{D-2}} \right)\right) dt^2 - \left({A^{ij}x^i \over r^{D-1}} + 
{\rm O}\left({1 \over r^{D-1}} \right) \right) dx^j dt \cr
&~~~ + \left[ \left(1 + {\mu \over r^{D-3}} + {\rm O}\left({1 \over
r^{D-2}} \right) \right) \delta_{ij} +{{e^{ij}} \over r^{D-3}} +
{\rm O}\left({1 \over r^{D-2}} \right) \right] dx^i dx^j.}}
$\mu$ is proportional to the ADM mass of the spacetime.  The anti-symmetric
tensor $A^{ij}$ is proportional to the angular momentum, and the symmetric
traceless tensor $e^{ij}$  represents gravitational radiation.  For an isolated
system, $\mu$ and $A^{ij}$ are constants.   In the case of string theory,
the above statements hold for the Einstein frame metric.  

We will  assume that we have a nonradiating system, and that a derivative
expansion is valid in the asymptotic region.  The latter assumption 
means that $\partial_\mu$ acting on any field brings down at least one
power of $1/r$.    The case with radiation present would be interesting
to study further, but is much more involved.

In the asymptotic region we will write 
$G_{\mu\nu} = \eta_{\mu\nu} + h_{\mu\nu}$  and expand the action \aa\  as
\eqn\ga{\eqalign{S & = 2\Phi(\infty)+{1 \over 2\pi \ap} \int \! d^2z \, \eta_{\mu\nu}
\p \Xt^\mu \bar{\p}\Xt^\nu \cr
& + {1 \over 2\pi \ap} \int \! d^2z \, \left\{ \left[
\p_\lambda h_{\mu\nu}(x) + {1 \over 3} H_{\mu\nu\lambda}(x)\right]
\p \Xt^\mu  \bar{\p}\Xt^\nu \Xt^\lambda
+ {e^{2\omega(z,\zb)}\over 4}\left[ \p_\lambda T_c(x) + \ap
R \p_\lambda \Phi(x) \right] \Xt^\lambda \right\} \cr
& + \ldots.}}

\subsec{Anti-symmetric tensor}

For the anti-symmetric tensor consider the operator
\eqn\gb{ {\cal O} = \left[ \p_\mu \Lambda_\nu(X) - \p_\nu \Lambda_\mu(X)
\right] \p \Xt^\mu \bar{\p}\Xt^\nu=  \p \left[\Lambda_\mu(X) \bar{\p}
\Xt^\mu  \right] - \bar{\p} \left[\Lambda_\mu(X) \p
\Xt^\mu  \right]. 
}
Since ${\cal O}$ is a total derivative, it can be added to the worldsheet
action without effect; this is a gauge transformation of $B_{\mu\nu}$.  
To derive the corresponding conserved charge we 
will use the fact that $\langle {\cal O} \rangle_{S_2}$ 
vanishes when integrated over the worldsheet. 

Now, in the free theory defined by the first line of \ga\ ${\cal O}$ is
 a dimension $(1,1)$ operator and we can define the product
of $X$'s by the standard normal ordering procedure. However, both of 
these statements are modified in the interacting theory with nontrivial
spacetime fields.  Fortunately, we are only interested in the 
structure of ${\cal O}$ in the asymptotic region, 
and this can be deduced by including the terms in the second line of \ga.

In particular, when we compute correlation functions with insertions of
${\cal O}$ we will find divergences from collisions with 
$\p \Xt^\mu  \bar{\p}\Xt^\nu \Xt^\lambda$.   Using the OPE to compute
the required counterterm we find the following 
renormalized operator in the asymptotic region (restricting attention
to the asymptotic region means we only keep terms with at most two spacetime
derivatives)
\eqn\gc{\eqalign{
 {\cal O} & = \p \left[ \Lambda_\mu(X) \bar{\p} \Xt^\mu - c\ln \Lambda
\left(\p_\lambda h_{\mu\nu}(x) + {1 \over 3} H_{\mu\nu\lambda}(x)\right)
\p^\mu \Lambda^\lambda(x) \bar{\p}\Xt^\nu \right] \cr
& - \bar{\p} \left[ \Lambda_\mu(X) {\p} \Xt^\mu - c\ln \Lambda
\left(\p_\lambda h_{\mu\nu}(x) + {1 \over 3} H_{\mu\nu\lambda}(x)\right)
\p^\nu \Lambda^\lambda(x) {\p}\Xt^\mu \right].}}

Now we use \bg.  We need only compute to first order in spacetime
fields to get the nonvanishing surface integrals.  So consider
\eqn\gd{ \int \! d^2z'\, (z'-z) e^{2\omega(z',\zb')} n_i
(\p_\mu \Lambda_\nu(x) - \p_\nu \Lambda_\mu(x)) 
\langle \p X^i (z',\zb') \p X^\mu(z,\zb)\bar{\p}
 X^\nu(z,\zb) \delta {\cal L}(z'',
\zb'') \rangle'_{0,S_2},}
where $\delta {\cal L}(z'',\zb'')$ stands for the operators appearing
in the second line of \ga.   If we choose a homogeneous metric on 
$S_2$ then $\langle {\cal O}(z,\zb) \rangle_{S_2}$ will be independent
of $z$; it is convenient to take $z=0$.  We will take $\Lambda_\mu$ to 
depend only on $X^0$, since this is the dependence that is needed in
order to derive conserved charges.  Now, $\p X^i (z',\zb')$ can 
contract against either $\p X^\mu(0)$, $\bar{\p} X^\nu(0)$ or a field in  
$\delta {\cal L}(z'',\zb'')$.  But the former case gives zero after
performing the angular part of the $d^2z'$ integral.   Furthermore,
$\p X^i (z',\zb')$ cannot contract against
the tachyon and dilaton terms in $\delta {\cal L}(z'',\zb'')$,
since this would leave a contraction between $\p X^\mu(0)$ and
$\bar{\p} X^\nu(0)$ which vanishes when anti-symmetrized.  This just 
leaves the contraction of $\p X^i (z',\zb')$ with the $h_{\mu\nu}$ and
$H_{\mu\nu\lambda}$ terms.  Performing the various contractions and
integrals, and cancelling divergences against the counterterm in \gc,
we find
\eqn\ge{ \langle {\cal O} \rangle_{S_2} = \int \! dt d^{D-2}S^i \,e^{-2\Phi}
\p_0 \Lambda_j
\left\{ a_1 H_{0ij} + a_2 \p_0 h_{ij} + a_3 \p_j h_{0i}
\right\}.}
The coefficients $a_{1,2,3}$ can in principle be computed, 
but this is not needed.
Assuming our standard falloff behavior the second and third terms do not
contribute to the surface integral.  The $a_2$ term 
vanishes because $\mu$ in \ha\ will be shown in the next subsection
to be conserved, and since $\p_0$ brings down at least one power of 
$1/r$ all other terms give vanishing surface integrals.  The expansion
\ha\ also immediately shows that the $a_3$ term does not contribute. 

  Now,  the left hand side of \ge\ vanishes upon integration over the
worldsheet since ${\cal O}$ is a total derivative, 
so upon integrating by parts on the 
right hand side we find the conserved charges
\eqn\gf{Q_j ~ \propto ~ \int \! d^{D-2}S^i \, e^{-2 \Phi}H_{0ij}.}
This agrees with the expected result from low energy field theory,
\elgc.  
A  typical situation is  to have translation invariance along some 
spatial direction $x^j$, giving the conserved charge per unit length
$\int \! d^{D-3}S^i \, H_{0ij}$.

\subsec{Gravitational field: ADM mass}

Our goal here is to derive a conserved energy-momentum vector
 from string theory.
More generally, we could also try to derive the conserved 
angular momentum tensor, 
include the effects of radiation, and so on.   Our scope
will be more limited, hopefully laying the groundwork for a more complete
treatment in the future.  Also, in string theory there are 
complications due to the dilaton which we do not entirely understand,
as will be discussed.  

The relevant symmetry to be considered is spacetime diffeomorphism 
invariance.  In particular, $\mu$ is the conserved  charge associated
with asymptotic time translation invariance. Let us then examine 
spacetime diffeomorphism invariance at the level of the sigma model
action \aa, (considering only the spherical worldsheet).  Consider the
variations 
\eqn\hb{\eqalign{\delta_\xi G_{\mu\nu}(X) & 
= \nabla_\mu \xi_\nu(X) + \nabla_\nu 
\xi _\mu(X), \cr
\delta_\xi B_{\mu\nu}(X) & = \xi^\lambda(X) \nabla_\lambda B_{\mu\nu}(X)
+ \nabla_\mu \xi^\lambda B_{\lambda \nu}(X) + \nabla_\nu \xi^\lambda
B_{\mu \lambda }(X), \cr
\delta_\xi T_c(X)  & = \xi^\lambda(X) \p_\lambda T_c(X), \cr
\delta_\xi \Phi(X) &  = \xi^\lambda(X) \p_\lambda \Phi(X).}}
The action \aa\ is invariant under the combination of \hb\  and
\eqn\hc{\delta_\xi X^\mu = - \xi^\mu(X).}
Therefore, the partition function
\eqn\hd{Z_{S_2}\left[G,B,T,\Phi\right] = \langle 1 \rangle_{S_2}
= \int \! {\cal D}X \, e^{-S}}
obeys
\eqn\he{ \delta_\xi Z_{S_2}\left[G,B,T,\Phi\right]=
V_M \int \! {\cal D}X \, \xi^\mu(X) {\delta S \over \delta X} e^{-S}
=- (V_M)^{D/2} \int \! d^Dx \, {\p \over \p x^\mu} \langle
\xi^\mu(X) \rangle'_{S_2}.}
In arriving at the last line of \he\ we have used the same chain
of manipulations as in \aj, and singularities in operator products have been
removed as in \an.
  \he\ is the expected behavior under a 
spacetime diffeomorphism.  A generic diffeomorphism invariant functional
\eqn\hf{ S =  \int \! d^Dx \, (-G)^{1/2} {\cal L}}
transforms under \hb\ as
\eqn\hg{\delta_\xi S = \int \! d^Dx \, (-G)^{1/2}\nabla_\mu \left\{
\xi^\mu {\cal L} \right\}
= \int \! d^Dx \,{\p \over \p x^\mu} \left\{
\xi^\mu (-G)^{1/2}{\cal L} \right\}
.}
\he\ and \hg\ are consistent when we remember that $\langle \xi^\mu 
\rangle'$ transforms as $(-G)^{1/2}$ times a vector.

As usual, we will consider a boundary at spatial infinity.  To obtain
true  symmetries of the partition function we can take $\xi^\mu$ tangent to 
the boundary,   $n_\mu \xi^\mu = 0$ so that $\delta_\xi Z =0$.  The conserved 
charges associated with these symmetries are mass, linear momentum, and
angular momentum.  

To show this, consider a Weyl invariant sigma model corresponding to some
asymptotically flat field configuration.  We take the matter fields to
fall off sufficiently rapidly at infinity so that no energy-momentum
flows through the boundary (we will make this more precise momentarily).
As  with the gauge field and anti-symmetric tensor,  the idea 
is to consider some on-shell variation of the gravitational field so
that the variation of the partition function is a boundary term.
We will set the tachyon to zero in  the following.  With the variations
in \hb\ define
\eqn\hga{ {\cal O} = \delta_\xi G_{\mu\nu}(X) \p X^\mu \bar{\p}X^\nu + 
\delta_\xi B_{\mu\nu}(X)\p X^\mu \bar{\p}X^\nu
+ {\ap \over 4} e^{2\omega} R \delta_\xi \Phi(X).}
Counterterms need to be added to define \hga\ as in \gc, but we will
not write them explicitly.  
Now, adding ${\cal O}$ to the worldsheet Lagrangian preserves Weyl
invariance, since its effect can be undone by the field redefinition
\hc.  Therefore, ${\cal O}$ must be a dimension $(1,1)$ operator, plus 
a total derivative on the worldsheet, plus a possible number times 
the Euler number density.  The latter is absent, as seen by taking 
$\xi^\mu =0$.  We know from \he\ 
(taking $n_\mu \xi^\mu =0$ at the boundary) that 
$\int \! d^2z\, \langle {\cal O}(z,\zb)\rangle_{S_2} = 0$.  Combining this
with \bg\ we have
\eqn\hgc{\int \! d^{D-1}S_i  \int \! d^2z \int \! d^2z'\, (z'-z) 
\langle\p X^i(z',\bar{z}') {\cal O}(z,\zb)\rangle'_{S_2} = 0.}
Which terms in \hga\ can contribute to the surface integral?  Some 
contributions correspond to the time rate of change of the gravitational
``charge'' and others correspond to currents flowing through the boundary.
The former are linear in the perturbations about flat space near the
boundary, while the latter are at least quadratic and so give vanishing
surface integrals provided one adopts standard falloff conditions.  It
is not hard to see that only the metric and dilaton can contribute linear
terms.  Therefore, we have
\eqn\hgd{\eqalign{\int \!& d^{D-1}S_i  \int \! d^2z \int \! d^2z'\, (z'-z)\cr
& \Big\langle\p X^i(z',\bar{z}')
\left\{ (\nabla_\mu \xi_\nu(X) + \nabla_\nu \xi_\mu(X))
\p X^\mu \bar{\p}X^\nu + 
{\ap \over 4} e^{2\omega(z,\zb)} R\xi^\mu \nabla_\mu \Phi(X)\right\}(z,\zb)
\Big\rangle'_{S_2} = 0.}}

At the boundary we can replace $\xi^\mu(X)$ by its constant mode part 
$\xi^\mu(x)$.   We  take $n^\nu \nabla_\nu \xi^\mu(x)=0 $ at the boundary
so that we can integrate by parts to get
\eqn\hl{\eqalign{ \int \! d^{D-1}S_i\, & \xi_\mu(x) \nabla_\nu \left\{
\int \! d^2z \int \! d^2z'\, (z'-z)e^{2\omega(z',\bar{z}')} \right. \cr
& \left. \Big\langle
\p X^i(z',\bar{z}')\left\{
\p X^\mu \bar{\p}X^\nu + \p X^\mu \bar{\p}X^\nu
-{\ap \over 4} e^{2\omega(z,\zb)}R \tilde{\Phi}(X)\eta^{\mu\nu}
 \right\}(z,\zb)\Big\rangle'_{S_2}\right\} = 0.}}
We defined $\tilde{\Phi} =\Phi - \Phi(\infty)$ since this is what 
contributes in \hgd, and replaced $G_{\mu\nu}$ by the asymptotic
Minkowski metric since deviations give vanishing surface integrals
when multiplied by $\tilde{\Phi}$.  
This implies the following expression for the conserved energy-momentum
\eqn\hm{\eqalign{ P^\mu ~ \propto ~ \int \! d^{D-2}S_i&
\int \! d^2z \int \! d^2z'\, (z'-z)e^{2\omega(z',\bar{z}')}\cr
& \Big\langle
\p X^i(z',\bar{z}')\left\{
\p X^0 \bar{\p}X^\mu + \p X^\mu \bar{\p}X^0
- {\ap \over 4} e^{2\omega(z,\zb)}R \tilde{\Phi}(X)\eta^{0\mu}
\right\}(z,\zb)
 \Big \rangle'_{S_2}.}}

We now check our result for the mass $P^0$ for the general asymptotically
flat background with vanishing tachyon.
  We work in coordinates such that the metric takes the form
\ha.  The anti-symmetric tensor and dilaton can  be taken to
fall off as $1/r^{D-3}$.  Since the dilaton adds some complications 
we will first consider the case in which the dilaton falls off faster
than  $1/r^{D-3}$, in which case we can disregard the dilaton term in \hm.

To evaluate the remaining correlation functions in 
\hm\ we need only use the first order perturbation of the sigma model 
around Minkowski space, since higher order terms will yield vanishing
surface integrals.  To compute $P^0$ we need to contract a $n_i \p X^i$,
a $\p X^0$, and a $\bar{\p}X^0$ against the first order perturbations in 
the sigma model action.  The contribution of the dilaton will be proportional
to  $n^i \p_i \p_0^2 \Phi(x)$
so we get a surface integral of a function falling off at least as
rapidly as $1/r^D$, and this vanishes.  Similarly, for the anti-symmetric
tensor, by anti-symmetry at least one of $\p X^0$ or $\bar{\p}X^0$ must
contract against an $X$ in $B_{\mu\nu}(X)$.  Then together with the
$\p X^i$ contraction we get the surface integral of a function falling
as least as rapidly as $1/r^{D-1}$, which vanishes.   This leaves only
the metric perturbation.   Examining the asymptotic form of the metric
given in \ha, we see that the only nonvanishing surface integral 
comes from an insertion of the operator $G_{00}(X) \p X^0 \bar{\p}X^0$.
Asymptotically,
\eqn\hn{G_{00}(X) \p X^0 \bar{\p}X^0 = -\left[1- {\mu \over r^{D-3}} + (D-3)
{\mu \over r^{D-1}} x^i \tilde{X}^i
 + {\rm O}\left({1 \over r^{D-1}}\right) \right]\p \tilde{X}^0 \bar{\p}
\tilde{X}^0.}
We therefore have
\eqn\ho{n_i\langle\p X^i(z',\bar{z}')\p X^0 \bar{\p}X^0(z,\zb)
 \rangle'_{S_2} ~ \propto ~
{\mu \over r^{D-2}}.}
Finally, from \hm\ we find
\eqn\hp{P^0 ~ \propto ~ \mu,}
which is the desired result.  There is a multiplicative factor relating
$\mu$ to the ADM mass and which is not fixed by our considerations since
we are just looking for a conserved quantity. 
 The numerical factor could  be
fixed by computing the explicit correlators in \hm.   
We chose to work in the center of mass coordinate system \ha\ in which 
$P^i = 0$, but we could  repeat the analysis in a boosted 
coordinate system.  The result for $P^\mu$ is of course fixed by
asymptotic Lorentz invariance, but it might be useful to check this
directly. 

Now we generalize by allowing the dilaton to fall off as 
$1/r^{D-3}$. From the low energy field theory analysis, we know 
that the conserved
energy-momentum is directly related to the asymptotic behavior of the
Einstein metric; see \emh.  In particular, if the $00$ component
of the string metric and dilaton behave as    
\eqn\hq{\eqalign{G_{00}(x) &~ \sim~ -1 + {\mu_s \over r^{D-3}} \cr
\Phi(x) & ~\sim ~\Phi(\infty) + {Q_\Phi \over r^{D-3}},}}
then, since the Einstein metric is 
$G^E_{\mu\nu} = e^{-4\tilde{\Phi}/(D-2)}G_{\mu\nu}$, the conserved 
energy is proportional to 
\eqn\hr{ \mu = \mu_s + {4 \over D-2} Q_\Phi.}
On the other hand, it is easy to verify that with the assumed fall off
conditions the formulas for momentum are unchanged by the presence of the
dilaton.

On the worldsheet, we see the corresponding  effect of the dilaton from \hm, 
which clearly  shifts the
conserved energy but not the momentum. Furthermore, the shift is proportional
to $Q_\Phi$, since $\langle \p X^i \Phi\rangle$ acts as a radial derivative
of $\Phi$, and then the surface integral picks out the leading 
 piece proportional to  $Q_\Phi$.  The ${4 \over D-2}$ prefactor in
\hr\ is more difficult to establish, as it involves 
computing relatively complicated correlators on a curved worldsheet
\foot{Part of the complication here is the lack of a convenient
regulator preserving the spacetime gauge symmetries, especially since
we need to work on a curved worldsheet for finiteness of $V_M$.  Dimensional
regularization suffers from ambiguities \TaniiUG\ due to the appearance of
$\lim_{d \rightarrow 2} (R_{ab} - {1\over 2}R g_{ab})/(d-2)$. 
Heat kernel regularization preserves the symmetries, but is 
awkward for extracting a finite part.}.
Presumably, the correct value of the prefactor follows from a general
worldsheet principle, but we have not been able to identify this so far.
This certainly deserves further study.

\newsec{Discussion}

The purpose of this work was to study the role of boundary terms
in solutions to string theory with noncompact target spaces.  The
particular application developed here was   defining conserved gauge 
charges as surface integrals at infinity.   In particular, this led 
to an intrinsic CFT definition of the energy-momentum of an
asymptotically flat spacetime.   Finding such a definition was an
outstanding challenge in early studies of string solitons (see, for 
example,  \CallanAT).  We have succeeded in doing this
here,  though a number of 
details such as the dilaton dependence  should certainly 
be developed more fully.  

There are several interesting open questions and directions for further
research.

\subsec{Anti-de Sitter Spacetimes}

In AdS, the conserved charges associated with diffeomorphisms are the
generators of conformal transformation of the boundary.   As shown by
Brown and Henneaux \BrownNW, this is especially interesting in the case of 
AdS$_3$, where one gets two copies of the Virasoro algebra with central charge
$c = 3 \ell/2 G$.   This result holds for any 
{\it asymptotically} AdS$_3$ spacetime, with the conserved charges 
expressed as surface integrals.  It is interesting to try to reproduce this
result from string theory, and for pure AdS$_3$ (with NS-NS B-field)
this was done in
\refs{\GiveonNS,\deBoerPP}.  This should generalize to the asymptotically
AdS$_3$ case using the approach developed here.  
If we write the metric in the form
\eqn\ia{ds^2 = d\phi^2 + g_{\mu\nu}(\phi, x)dx^\mu dx^\nu,}
the conserved energy-momentum tensor
can then be presumably be written as roughly (modulo dilaton terms)
\eqn\ib{ T^{\mu\nu} \sim 
\int \! d^2z \int \! d^2z'\, (z'-z)e^{2\omega(z',\bar{z}')}\langle
\p \phi(z',\bar{z}')
(\p X^\mu \bar{\p}X^\nu + \p X^\nu \bar{\p}X^\mu)(z,\zb)
 \rangle'_{S_2}.}

\subsec{Spacetime action and the string partition function}

Since in field theory conserved charges arise as symmetries of the
action, it would seem that the most efficient way to pass to string theory
would be to to first give a CFT definition of the spacetime action.  
The idea that  the spacetime action should be closely related to the 
string partition function goes back to work of Fradkin and Tseytlin 
\FradkinPQ\ and
has been investigated by many authors since, but is still not completely
understood.  Some of the confusions have to do with obtaining a
suitable off-shell action, which is perhaps unnatural in a theory with
gravity.  Fortunately, to derive conserved charges one only needs the 
on-shell action, and it seems reasonable to suppose that this is equal
to the string theory vacuum-to-vacuum
 amplitude.  For the open string at the level of
disk amplitudes this works nicely (and can be extended off-shell)
\refs{\FradkinQD,\WittenQY, \GerasimovZP,\KutasovQP}:
\eqn\ic{S_{{\rm open}} = {1 \over {\rm Vol}_{SL(2,R)}} 
\left( {\rm det'}_{D_2} P_1^\dagger P_1 \right)^{1/2} e^{-\lambda}Z_{D_2}.}
The functional integrals can be computed unambiguously
\refs{\DouglasEU,\WeisbergerQD}, $e^{-\lambda}$
is related to the gravitational coupling  $\kappa$ by unitarity \Polch, 
and ${\rm Vol}_{SL(2,R)}$ can be
assigned a finite ``renormalized'' value \LiuNZ.  For instance, the correct
value of the D-brane tension can be obtained this way.   So for the 
open string we could have formulated things in this framework.

For the closed string the formula  analogous to \ic\ is not as successful.
The main problem is that ${\rm Vol}_{SL(2,C)}$ is divergent even after
``renormalization'' due to the appearance of a logarithmic divergence.  
Tseytlin \TseytlinTV\ 
has proposed a spacetime action by essentially cancelling this
divergence against a similar divergence in the sphere partition function. 
Unfortunately, this proposal does not seem to successfully reproduce
the boundary terms in the spacetime action \KazakovPJ.  The latter are crucial,
especially in our context; for instance they give the entire result in
spacetimes with constant dilaton.    

A better understanding of this issue is important 
for the AdS/CFT correspondence,
since this is supposed to equate the boundary CFT partition function 
in the presence of
sources to the spacetime action with prescribed boundary conditions. 
The former is generically nonzero, but the latter naively vanishes
in string theory due to the division by ${\rm Vol}_{SL(2,C)}$.  It
has been suggested \GiveonNS\ that for AdS$_3$ there is a compensating 
divergence from the spacetime volume integration, given that Euclidean 
AdS$_3$ is the coset $SL(2,C)/SU(2)$.
Besides the fact that this cancellation is specific to  AdS$_3$,
it does not give the correct result even in this case.  A direct 
comparison can be made since the sphere partition
function for AdS$_3$ is readily computed, and the partition function 
of the boundary theory (in the absence of sources but for a general 
boundary metric) is determined by the conformal anomaly.

We believe that the resolution is along the lines of the present paper.
We have seen that in noncompact spacetimes worldsheet $SL(2,C)$ symmetry
can be violated by boundary terms, so simply dividing all amplitudes
by the divergent ${\rm Vol}_{SL(2,C)}$ factor is not justified. 
Hopefully, the correct procedure leaves boundary terms in a way
similar  to what
we have seen here for one-point functions.

\bigskip\medskip\noindent
{\bf Acknowledgements:} We thank Eric D'Hoker, David Kutasov, Emil
Martinec, and Arvind Rajaraman for helpful discussions.
This work was supported  by NSF grant PHY-0099590.

\listrefs

\end